\documentclass[preprint,12pt]{elsarticle}

\usepackage{graphicx}
\usepackage{epstopdf}

\usepackage{amssymb}

\journal{Journal of Physics G}

\begin{document}

\begin{frontmatter}



\title{Extracting the hadronization timescale in $\sqrt{s}=7$ TeV proton-proton collisions
from pion and kaon femtoscopy}


\author{T. J. Humanic}

\address{Department of Physics, Ohio State University, Columbus, OH, USA}

\begin{abstract}
A hadronic rescattering model with the proper hadronization time width as the free parameter
is compared with charged pion and charged and neutral kaon femtoscopy measurements
from the LHC ALICE experiment for $\sqrt{s}=7$ TeV proton-proton collisions. Comparisons between the
model and measurements are made for one-dimensional source parameters in several
charged multiplicity and transverse particle pair momentum bins. It is found that a reasonable
description of the measured source parameters by the model is obtained for a hadronization proper
time width of 0.4 $\pm$ 0.1 fm/c, which is in agreement with an estimate based on the uncertainty
principle. The model calculations also suggest that 1) some form of collectivity is
necessary to describe the multiplicity dependence of the measured radius parameters, and
2) the underlying physical size and timescale of the collision is significantly larger than what the
extracted radius parameters and hadronization proper time width would imply.

\end{abstract}



\end{frontmatter}


\section{Introduction}
Of fundamental interest in studying high-energy particle collisions is how hadrons are formed
in these collisions. An important element in fully understanding the hadronization process is 
knowledge of the
space-time position where the hadronization takes place during the collision. In proton-proton collisions, the spacial
position of hadronization transverse to the collision axis is characterized by the radius of a proton, i.e. 1 fm,
whereas the hadronization timescale and the longitudinal spacial position, which depends on 
the product of the longitudinal velocity and hadronization timescale, are not well defined.
An order-of-magnitude estimate for the hadronization proper time of a particle of rest mass $m_0$ can 
be obtained by applying the uncertainty principle in the rest frame of the particle, 
\begin{eqnarray}
\Delta \tau\approx\frac{\hbar}{2\Delta E}
\label{dtau1}
\end{eqnarray}
where $\Delta E$ is the uncertainty in the particle energy and $\Delta \tau$ is the uncertainty
in the particle hadronization proper time, i.e. the width of the hadronization proper time
distribution.
Since Eq. \ref{dtau1} is evaluated in the rest frame of the particle, then $\Delta E=\Delta m_0 \sim m_0$,
where $\Delta m_0$ has been set to $m_0$ as an order-of-magnitude estimate. Thus, Eq. \ref{dtau1} becomes,
\begin{eqnarray}
\Delta \tau \approx\frac{\hbar}{2m_0}\leq 0.7\ {\rm fm/c}
\label{dtau2}
\end{eqnarray}
where, as the lightest hadron,  the pion rest mass sets the upper limit on the width of the hadronization proper time distribution.

Several theoretical \cite{csorgo,chmaj} and phenomenological \cite{hum1,truesdale,Humanic:2006ib} 
studies of the hadronization (or formation) time in proton-proton collisions 
have been made over the past 30 years. The phenomenological ones have attempted to
extract a single hadronization proper time in each case by comparing a model with two-pion femtoscopy measurements
for $\sqrt{s}=$ 0.2 \cite{hum1}, 0.9 \cite{truesdale}, and 1.8 \cite{Humanic:2006ib} TeV collisions, obtaining
values in the range of 0.1-0.2 fm/c. In a recent phenomenological study comparing the UrQMD 
model \cite{urqmd} with
LHC ALICE collaboration 3-dimensional two-pion femtoscopy measurements in $\sqrt{s}=7$ TeV
proton-proton collisions \cite{Aamodt:2011kd} a range for the single hadronization proper time 
of 0.3-0.8 fm/c was obtained. Although the extracted values in these studies are for $\delta$-function distributions
of the hadronization proper time rather than widths of finite distributions, the extracted values seem consistent 
with the order-of-magnitude limit given in Eq. \ref{dtau2}.

The main goal of the present work is to extract hadronization time information by comparing
a rescattering model similar to that used in Refs. \cite{hum1,truesdale,Humanic:2006ib}  with LHC ALICE 1-dimensional
$\pi^{ch}\pi^{ch}$, K$^{ch}$K$^{ch}$, and K$^0_s$K$^0_s$ femtoscopy measurements in $\sqrt{s}=7$ TeV
proton-proton collisions \cite{Aamodt:2011kd,kch,k0s}. The main differences between the present work
and the previous ones are the following:

\begin{itemize}
\item Rather than using a single hadronization proper time, a more physically motivated 
Gaussian distribution of proper times characterized by a width, $\tau_{had}$, is used in 
the present work, as in Ref. \cite{csorgo}. $\tau_{had}$ will be more directly comparable with
$\Delta\tau$ in Eq. \ref{dtau2}.

\item The present rescattering model uses identified charges and anti-particles, rather than 
being isospin averaged as in the previous works, and also includes $p\bar{p}$ and $n\bar{n}$
annihilation,

\item Rather than only making model comparisons with two-pion femtoscopy measurements, 
the present work will also compare with charged and neutral kaon measurements. This should serve
to place additional constraint on the extracted hadronization proper time information.
\end{itemize}

The remainder of the paper is organized into a description of the model, results and discussion, and
the summary.

\section{Description of the model}
The rescattering model used is similar to that used in Refs. \cite{hum1,truesdale,Humanic:2006ib}
and is briefly described here.
The PYTHIA version 6.4 code \cite{pythia} is used to generate the initial particles and their momenta for
$\sqrt{s}=7$ TeV proton-proton collisions. This version of PYTHIA has been found by the LHC
ATLAS \cite{atlas} and CMS \cite{cms1,cms2} collaborations to describe the measured pion and charged 
and neutral kaon transverse momentum distributions from $\sqrt{s}=7$ TeV proton-proton collisions 
reasonably well. Also, the ALICE collaboration
has used a version of PYTHIA to model the correlation function backgrounds in its pion and
charged and neutral kaon femtoscopy measurements in $\sqrt{s}=7$ TeV 
proton-proton collisions \cite{Aamodt:2011kd,kch,k0s}. The particles extracted from PYTHIA and
input into the model, including all charge states and anti-particles, are $p$,$n$,$\pi$,$K$,$K^*$,$\Lambda$,
$\Delta$, $\rho$, $\omega$, $\eta$, $\eta'$and $\phi$. The allowed two-body scattering processes
and decays, which are constrained to obey conservation of energy, momentum, charge, baryon
and strangeness number, are:
\begin{itemize}
\item elastic ($X$ and $Y$ represent any particles)
\begin{eqnarray}
XY\rightarrow XY\nonumber
\end{eqnarray}
\item inelastic and decays ($N$ here represents a nucleon or anti-nucleon)
\begin{eqnarray}
\pi N\leftrightharpoons \Delta, \pi\Delta, K\Lambda \ \ \ \ \ \ NN\leftrightharpoons N\Delta \ \ \ \ \ \ KN\leftrightharpoons \pi\Lambda\nonumber \\
\pi K\leftrightharpoons K^* \ \ \ \ \ \ \pi\pi \leftrightharpoons \rho, \eta \ \ \ \ \ \ \pi\pi\rightarrow\phi\leftrightharpoons KK\nonumber \\
\pi\pi\rightarrow \omega, \eta' \rightarrow \pi\pi\pi \ \ \ \ \ \ \Lambda\rightarrow \pi N\nonumber
\end{eqnarray}
\item $N\bar{N}$ annihilation
\begin{eqnarray}
N\bar{N} \rightarrow \rho\rho, \rho\omega, \omega\omega, \rho\eta, \omega\eta, \eta\eta
\end{eqnarray}
\end{itemize}
The annihilation cross sections were taken from Ref.~\cite{annih1} and branching ratios from Ref.~\cite{annih2}. 

The space-time point of the $i^{th}$ particle of rest mass $m_{0i}$ at hadronization in the proton-proton collision frame 
$(x_i,y_i,z_i,t_i)$ with PYTHIA energy-momentum $(p_{xi},p_{yi},p_{zi},E_i)$ is determined in the model by a Gaussian distribution
for the transverse radius, $r_T$, and, as mentioned earlier,  a Gaussian distribution for the 
proper time, $\tau$, such that
\begin{eqnarray}
\frac{dn}{dr_{Ti}}\propto\exp(-\frac{r_{Ti}^2}{2\sigma_r^2}) \ \ \ \ \ \ \ \ \
\frac{dn}{d\tau_i}\propto\exp(-\frac{\tau_i^2}{2\tau_{had}^2})
\label{prob}
\end{eqnarray}
\begin{eqnarray}
x_i=r_{Ti}\cos\phi_i \ \ \ \ \ y_i=r_{Ti}\sin\phi_i \ \ \ \ \ z_i=\tau_i\frac{p_{zi}}{m_{0i}} \ \ \ \ \ t_i=\tau_i\frac{E_i}{m_{0i}}
\label{geom}
\end{eqnarray}
where $\sigma_r$ is set to 1 fm, $\phi_i$ is the azimuthal angle of the $i^{th}$ particle set randomly between
$0-2\pi$, and $\tau_{had}$ is the hadronization proper time width, which is a free parameter to be adjusted
to get the best agreement with measurements.

Each particle is allowed to evolve from these initial conditions in time steps of 0.25 fm/c undergoing 
scatterings and decays until no more scatterings or decays occur. At this point, freeze-out of the particle
occurs and after this point the particle is free streaming.

Quantum statistics is imposed pair-wise on boson pairs $a$ and $b$ by weighting them at their 
freeze-out phase-space points $(\overrightarrow{r_a},t_a,\overrightarrow{p_a},E_a)$ and
$(\overrightarrow{r_b},t_b,\overrightarrow{p_b},E_b)$ with
\begin{eqnarray}
W_{ab}=1+\cos(\Delta\overrightarrow{r}\cdot\Delta\overrightarrow{p}-\Delta t\Delta E)
\end{eqnarray}
where,
\begin{eqnarray}
\Delta\overrightarrow{r}=\overrightarrow{r_a}-\overrightarrow{r_b}\ \ \ \ \
\Delta\overrightarrow{p}=\overrightarrow{p_a}-\overrightarrow{p_b}\ \ \ \ \
\Delta t=t_a-t_b\ \ \ \ \Delta E=E_a-E_b
\end{eqnarray}
The correlation function, $C(q_{inv})$, is formed by binning pairs in terms of the invariant momentum difference
$q_{inv}=\left | \Delta\overrightarrow{p}\right | -\left | \Delta E\right |$ as the ratio of
weighted pairs, $N(q_{inv})$, to unweighted pairs, $D(q_{inv})$,
\begin{eqnarray}
C(q_{inv})=\frac{N(q_{inv})}{D(q_{inv})}
\label{cf_model}
\end{eqnarray}
Since there are no final-state interactions in the model between boson pairs after freeze out such as 
Coulomb or strong interactions, a simple Gaussian function is fitted to Eq.\ref{cf_model} to extract the
boson source parameters which are compared with experiment,
\begin{eqnarray}
C_{\rm fit}(q_{\rm inv})=\alpha[1+\lambda\exp(-q_{\rm inv}^2R^2)]
\label{cf_fit}
\end{eqnarray}
where $R$ is the radius parameter which, in principle, is related to the size of the boson source,
$\lambda$ is a parameter that reflects the strength of the quantum statistics effect as well as the degree
to which the Gaussian function fits to the correlation function, and $\alpha$ is an overall normalization
parameter. For the full effect of quantum statistics and a perfect Gaussian fit to the correlation function,
$\lambda$ has the value of unity. Eq. \ref{cf_fit} is the same function as used by ALICE to extract $R$
and $\lambda$ from quantum statistics in their measurements.

Cuts on charged particle multiplicity ($N_{ch}$), transverse momentum ($p_T$), pseudorapidity ($\eta$),
pair transverse momentum ($k_T=\left | \overrightarrow{p_a}+\overrightarrow{p_b}\right | /2$) and $q_{inv}$ are applied to the model particles
to duplicate those made by ALICE in their measurements. Also, as done by ALICE, like-charge pairs
of both signs are summed for pions and charged kaons, and for neutral kaons all pair combinations of
$K^0$ and $\overline{K^0}$ are summed.

\section{Results and Discussion}
\begin{figure}
\begin{center}
\includegraphics[width=135mm]{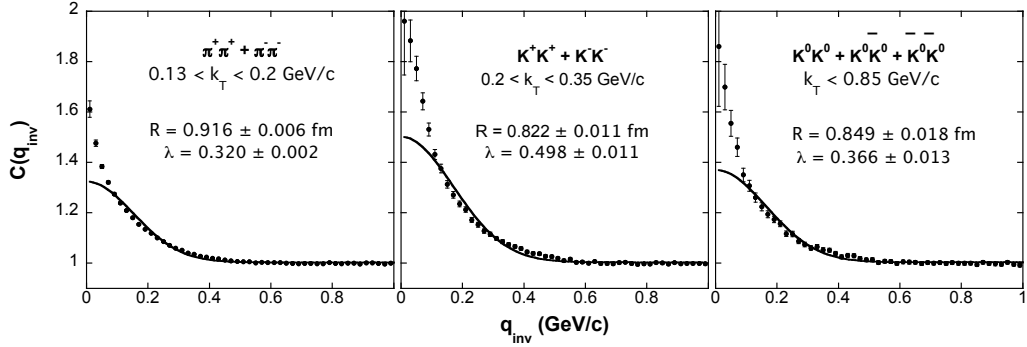} \caption{Sample correlation functions from $\sqrt{s}=7$ TeV
proton-proton collisions for charged pions,
charged kaons, and neutral kaons from the model using Eq. \ref{cf_model}. $N_{ch}$ is
in the range 1-11 and  $\tau_{had}=0.4$ fm/c . Fits of Eq. \ref{cf_fit} to the 
model points are also shown.}
\label{fig1}
\end{center}
\end{figure}

Figure \ref{fig1} shows sample correlation functions from $\sqrt{s}=7$ TeV proton-proton collisions 
for charged pions, charged kaons, and neutral kaons from the model calculated from Eq. \ref{cf_model}. The
range in $N_{ch}$ is 1-11 and $\tau_{had}=0.4$ fm/c. Fits of Eq. \ref{cf_fit} to the model points 
are also shown. As seen, $R\sim 1$ fm, as would be expected for proton-proton collisions,
and $\lambda$ is significantly less than the idealized case of unity. In all three plots, the correlation
function is seen to be more exponential in shape than Gaussian which has a significant effect on
lowering $\lambda$. In addition, for the pion plot, the intercept of the correlation function is itself
less than unity due to the effect of the long-lived resonances $\eta$ and $\eta'$ diluting the
strength of the correlation function.

\begin{figure}
\begin{center}
\includegraphics[width=135mm]{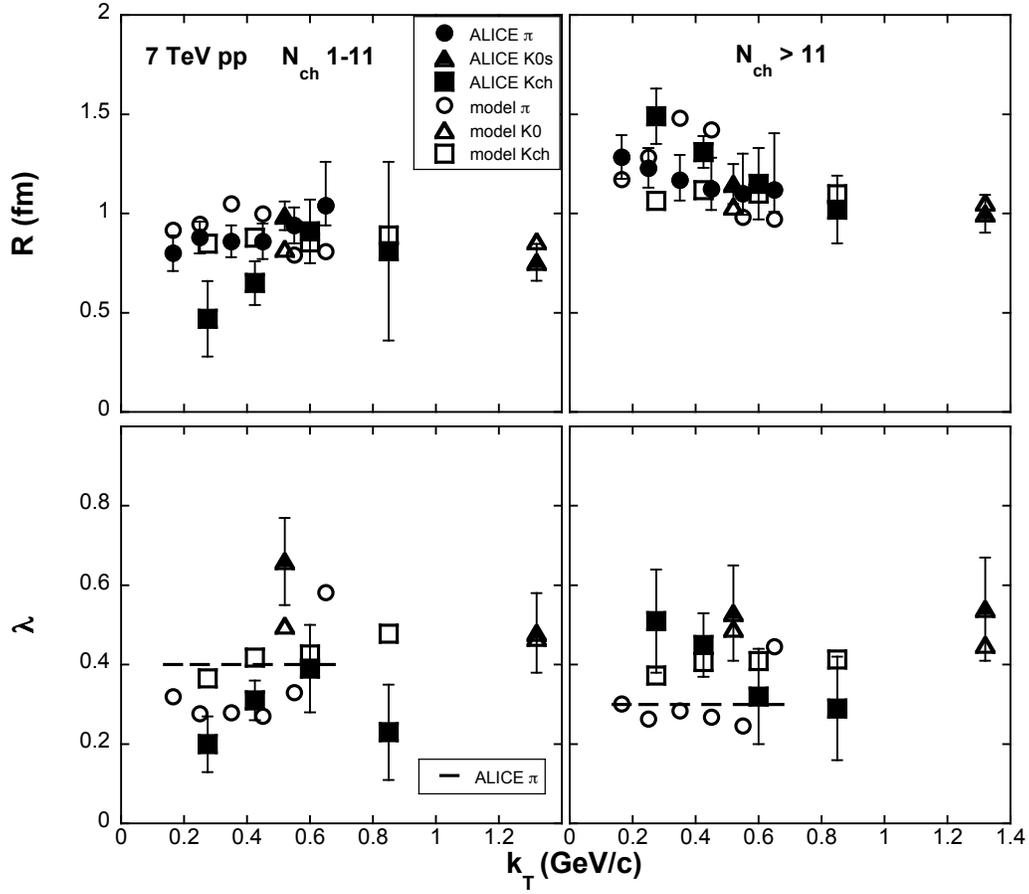} \caption{Comparison of source parameters from
the model using $\tau_{had}=0.4$ fm/c with ALICE measurements in two charged particle
multiplicity ranges.}
\label{fig2}
\end{center}
\end{figure}

Figure \ref{fig2} shows a comparison of the model source parameters using $\tau_{had}=0.4$ fm/c with ALICE measurements in two charged particle multiplicity ranges: $N_{ch}$ 1-11 and 
$N_{ch}>11 $ \cite{Aamodt:2011kd,kch,k0s}. The error bars are statistical+systematic for the ALICE
points and are statistical but equal to or less than the marker size for the model points. The approximate scales
of the ALICE $\lambda$ parameters for pions are indicated by the dashed lines \cite{adam}, since the specific
values were not published by ALICE in Ref. \cite{Aamodt:2011kd}. Although there are specific differences
seen between the model and ALICE, overall 
the model points describe the
gross features of the ALICE measurements for pions and charged and neutral kaons 
reasonably well for both R and $\lambda$: 1) the overall scales of R for the different particle types 
agree with ALICE,  2) the model reproduces the increase in R in going from lower to higher $N_{ch}$ 
as measured by ALICE, and 3) the model reproduces the scales of the measured $\lambda$ parameters.
The dependence on $\tau_{had}$ of the agreement of the model with ALICE was studied and it was
found that $\tau_{had}=0.4$ fm/c gave the best agreement. For $\tau_{had}<0.4$ fm/c the initial density
of particles of the system is higher, as suggested by Eqs. \ref{prob} and \ref{geom}, resulting in more
rescattering and thus larger R values, whereas for $\tau_{had}>0.4$ fm/c the initial density is lower but
the overall timescale is higher again producing larger R values. From this study the uncertainty 
in $\tau_{had}$ is estimated, giving the best value $\tau_{had}=0.4\pm0.1$ fm/c. This value is in
agreement with the upper limit estimate given in Eq. \ref{dtau2}.

\begin{figure}
\begin{center}
\includegraphics[width=135mm]{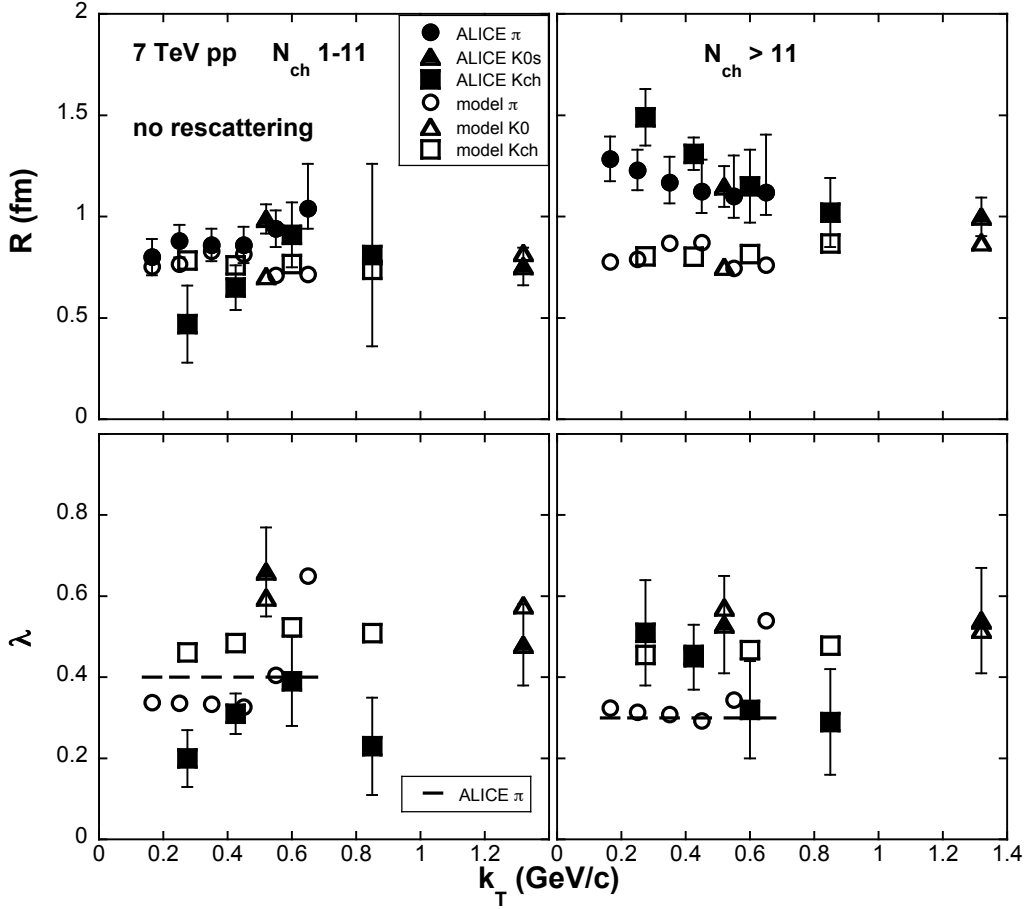} \caption{Same as Figure \ref{fig2} but with rescattering turned
off in the model.}
\label{fig3}
\end{center}
\end{figure}

It is interesting to study the effect that rescattering has on the model source parameters. Figure \ref{fig3}
shows a similar comparison of the model with ALICE as shown in
Figure \ref{fig2} except that rescattering is turned off in the model (decays are still allowed).
As seen, while the model agreement with ALICE for $\lambda$ is not affected much by
turning off rescattering, the model $R$
parameters are strongly affected in that they no longer depend on $N_{ch}$ and are significantly
smaller than ALICE for $N_{ch}>11$. This suggests that some kind of ``collective motion'' of the system,
in the present model being due to rescattering, is needed to explain the measured $N_{ch}$
behavior of $R$ in $\sqrt{s}=7$ TeV proton-proton collisions. The existence of collective
behavior in proton-proton collisions has also been suggested in other works
\cite{Bozek:2009dt,Werner:2011fd,Kisiel:2010xy,Chajecki:2009es}.

\begin{figure}
\begin{center}
\includegraphics[width=135mm]{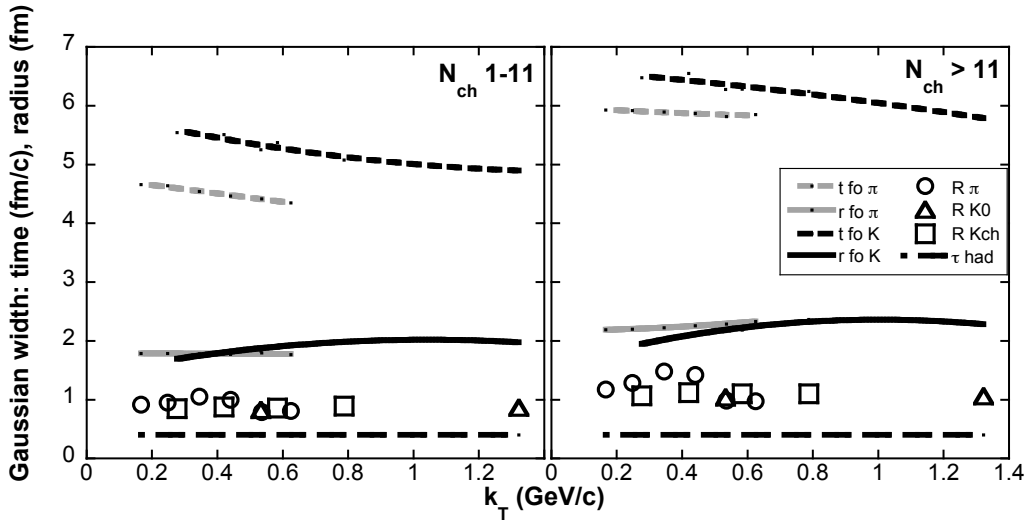} \caption{Model comparisons between $R$ and $\tau_{\rm had}$ and
Gaussian widths of the source radius and time distributions.}
\label{fig4}
\end{center}
\end{figure}

The model connections  between the extracted $R$ and $\tau_{had}$ values and the
underlying physical sizes and timescales in the collision frame are shown in Figure \ref{fig4}. The short-dashed
lines represent the equivalent Gaussian widths of the freeze-out time distributions
and the solid lines represent the equivalent Gaussian widths of the freeze-out radius distributions. 
The freeze-out time and radius width calculations are restricted to the ranges
$t_{FO}<20$ fm/c and $r_{FO}<20$ fm, respectively, to exclude the trivially large timescales and sizes which
would be introduced by long-lived resonances.
Pions (grey lines) and the combined kaons (black lines) are plotted separately.
The model $R$ parameters are the same as plotted in Figure \ref{fig2}. As seen, although the extracted
$R$ and $\tau_{had}$ are $\sim 1$ fm and 0.4 fm/c, respectively, the underlying physical sizes and timescales
are $\sim 2\times$ and $\sim 15\times$ larger, respectively. The large physical sizes and timescales
are due to a combination of time dilation in the initial state, resonance decays and rescattering dynamics. It is also seen that the increase
in the $R$ parameters in going from lower to higher $N_{ch}$ is also reflected in an increase in the
physical radii and timescales for pions and kaons, this effect being completely driven in the model by rescattering,
as already suggested in comparing Figures \ref{fig2} and \ref{fig3}.

\begin{figure}
\begin{center}
\includegraphics[width=80mm]{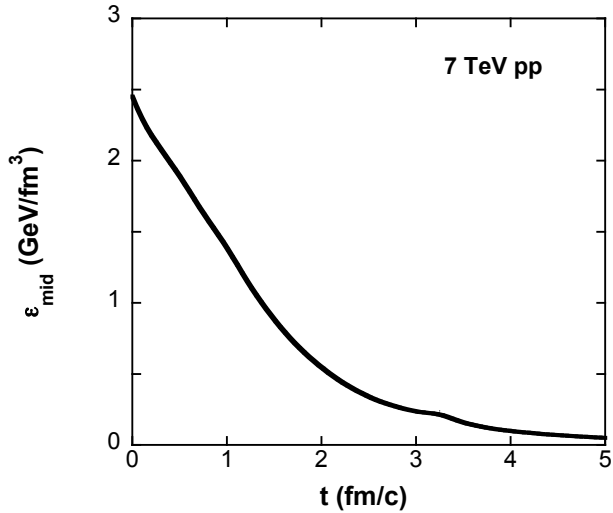} \caption{Time evolution of the energy density at midrapidty for the first
5 fm/c from the model averaged over all multiplicities for $\tau_{had}=0.4$ fm/c.}
\label{fig5}
\end{center}
\end{figure}

Figure~\ref{fig5} shows the time evolution of the energy density at midrapidty, $\epsilon_{mid}$, for the first 5 fm/c from the model averaged over all multiplicities for $\tau_{had}=0.4$ fm/c. Besides the expected 
decrease in $\epsilon_{mid}$ as time evolves due to the expansion of the system, it is seen that
$\epsilon_{mid}>1$ for $t<1.35$ fm/c, reaching $\epsilon_{mid}\approx2.4$ GeV/fm$^3$ at the initial time
in the collision. An estimate of the initial energy density in $\sqrt{s}=7$ TeV proton-proton collisions based on
hydrodynamic calculations has been carried out in Ref.~\cite{edpphydro}, and it is also found there that
energy densities greater than 1 GeV/fm$^3$ are initially produced in these collisions, suggesting that
a non-hadronic state may be present at the early stages, as is thought to be the case
in high-energy heavy-ion collisions. In this scenario, the collision evolves from an initial hydrodynamic state
made up of partons into the hadronic state with some possible rescattering and finally to freeze-out from 
which experimental observables are constructed. In this case, one would expect that collective effects 
seen in the observables would be due to some combination of hydrodynamics of the partons and 
final-state hadronic rescattering. The possibility of this hydrodynamic scenario leads to questioning
the validity of a purely hadronic picture as is used in the present work which serves as a
``limiting case'' picture. If a partonic state is indeed initially present, it must evolve eventually into hadrons, 
and this evolution would likely be gradual in time as opposed to a sudden hadronization
of the system as would happen in a first-order phase transition. The simple hadronic rescattering picture
could have some degree of validity during this mixed-phase transition period when 
perhaps ``quasi-hadrons'' are present. Even at the earliest times when the system may be purely
partonic and thus better described by a hydrodynamic picture, hadronic rescattering is perhaps
able to mimic to some degree the early hydrodynamic evolution. Thus, in this
partonic scenario the present purely
hadronic rescattering model might be thought of as mimicking a ``viscous'' hydrodynamic
evolution of the system.

\section{Summary}
In this work, a hadronic rescattering model with the proper hadronization time width as the free parameter
was compared with charged pion and charged and neutral kaon femtoscopy measurements
from the LHC ALICE experiment for $\sqrt{s}=7$ TeV proton-proton collisions. Comparisons between the
model and measurements were made for one-dimensional source parameters in several
charged multiplicity and transverse particle pair momentum bins. It is found that a reasonable
description of the measured source parameters by the model is obtained for a hadronization proper
time width of 0.4 $\pm$ 0.1 fm/c, which is in agreement with the estimated upper limit based on the uncertainty principle given in Eq. \ref{dtau2}. The model calculations also suggest that 1) some sort of collectivity is
necessary to describe the multiplicity dependence of the measured radius parameters, and
2) the underlying physical size and timescale of the collision is significantly larger than what the
extracted radius parameters and hadronization proper time width would suggest. It will be interesting to
repeat this study once new data are available from the upgraded LHC for $\sqrt{s}=14$ TeV proton-proton
collisions since the anticipated higher particle multiplicities produced in these collisions should enhance
the amount of collectivity present and thus enhance the magnitude of the effects studied in the present
work. It would also be of value in characterizing the early stages of these collisions to study penetrating
probes such as direct photons and dileptons \cite{edpphydro} to obtain direct information on the source 
of the collectivity suggested in these femtoscopy studies.





The author wishes to acknowledge financial support from the U.S.
National Science Foundation under grant PHY-1307188, and to acknowledge computing
support from the Ohio Supercomputing Center.






\end{document}